\documentclass{article}
\usepackage{a4wide,graphics,amsmath,amssymb}
\begin{document}


\title{Squeezed Condensates \\ \small{submitted to Phys. Rev. A}}

\author{Uffe V. Poulsen and Klaus M{\o}lmer
\\
 \small{Institute of Physics and Astronomy, University of Aarhus,
    DK-8000 \AA rhus C, Denmark}
\\\small{email: uvp{@}ifa.au.dk}
}

\maketitle

\begin{abstract}
  We analyse the atomic state obtained by photo-dissociation
  of a molecular Bose-Einstein-condensate. This process is equivalent
  to down-conversion in quantum optics where it is responsible for
  squeezing of the field amplitudes. Monte Carlo simulations derived
  from the Positive P description of the system, and approximate
  equations for second moments of the atomic field operators are
  introduced to describe the early stages of the process -- not
  amenable to the usual mean-field description.
\end{abstract}  


\section{Introduction}
\label{sec:introduction}

Like electromagnetic waves, atomic matter waves can to a large extent
be manipulated in space. Interference phenomena have been observed,
and a large number of high precision atom optics measurements have
been carried out~\cite{atom_optics}. One of the potential future
applications of atomic Bose-Einstein condensates is as coherent matter
wave sources for interferometric time and frequency standards,
detection of inertial effects, and a host of related technological
tasks.  The
spatial~\cite{andrews97:_obser_inter_between_two_bose_conden} and
temporal~\cite{hagley99:_measur_coher_bose_einst_conden} coherence of
condensates has been verified, and coherent amplification of matter
waves has been demonstrated~\cite{ketterle-gain} to establish the
close analogy with laser and maser sources of light.

For high precision purposes, however, it was realized long time ago,
that so-called non-classical states of light may be more useful than
classical field states~\cite{fabre92:_quant_noise_reduc_optic_system}.
It is therefore natural to consider the production of such
'non-classical' states of atoms as well. The term 'non-classical' is
ambiguous here, since what is very non-classical for light, {\it e.g.}
a number state, can seem perfectly classical for atoms, and we shall
instead use the term quantum correlated states. Due to the collisional
interaction between atoms in a Bose condensate, this system contains a
non-linearity, in equivalence with the Kerr-effect for light, and this
collisional interaction has already been observed \cite{deng99:_four}
in a matter wave analog of four-wave mixing in non-linear optics.
There have been several proposals to utilize collisional effects to
produce certain quantum correlated states, such as Schr\"odinger
cat-like states \cite{cat}, and to observe the effect
of the non-linearity on the dynamics of the condensate
\cite{tunneling}. It is clear from these studies that, even though the
collisional interaction can be controlled experimentally
\cite{tuning} it is
not easy to control precisely the quantum features of interest; the
problem is a genuine multi-mode one due to the spatial degrees of
freedom, and one has to pay attention to the role of non-condensed
atoms, as well.

In this Paper, we suggest to follow instead the procedure of quantum
optics for production of squeezed light: we suggest to implement an
'atomic OPO'. The optical parametric oscillator (OPO) is a device
where light is down converted, so that a single pump photon at
frequency $2\omega$ is converted into two photons, each of frequency
$\omega$ (degenerate case). The process is the inverse of second
harmonic generation (SHG), and in practical experiments, one often
sees a strong field of frequency $\omega$, which is first frequency
doubled in the SHG crystal, and the high frequency field is
subsequently down-converted in a similar non-linear crystal, now
working as an OPO.

The matter wave analogue of the SHG process is one in which the atoms
are combined by photo-association into di-atomic molecules. This
process has been analyzed
theoretically~\cite{mackie00:_bose_stimul_raman_adiab_passag_photoas},
and it was recently demonstrated experimentally, that part of an
atomic condensate may be converted by stimulated Raman transitions
into molecules this way \cite{wynar00:_molec_bose_einst_conden}. We
suggest to follow the analogy with the OPO quite directly, {\it i.e.},
to apply laser fields to drive the Raman transition from the molecular
state back to free atoms (photo-dissociation), and we note, that the
quantum correlations appear due to the fact that only even number
states for the atomic component will be present in the sample, because
atoms are created in pairs.

The dynamics of coupled atomic and molecular degenerate gasses has
been studied in some detail~\cite{molecules_theory} with particular focus on
the spatial and temporal dependence of the mean field dynamics, {\it
  e.g.}, the appearance of
solitons~\cite{drummond98:_coher_molec_solit_bose_einst_conden}. We
focus on the ``quantum optics'' aspects, {\it i.e.}, the atom number
distribution, and the consequences for atom counting experiments.

The organization of this paper is as follows: In Sec. \ref{sec:model},
we present our proposal, and we derive two theoretical methods, one
approximate and one exact, to analyze the dynamics of the system. In
Sec. \ref{sec:results} we present numerical results for various
relevant quantities. In Sec. \ref{sec:appli} we analyze the use of our
quantum correlated atoms as squeezed input in  two-state Rabi-oscillations
with a large, ``classical'', condensate. In Sec. \ref{sec:conclusion},
we discuss the results and we briefly indicate some alternative ideas
for the production and application of quantum correlated condensates.

\section{The model}
\label{sec:model}

\subsection{Trapped cold atoms in second quantization}
\label{sec:second_quan}

It is convenient to describe dilute gasses of cold atoms in a trap in
the second quantized formalism. In this description the Hamiltonian
for the system at low temperatures is usually taken to be:
\begin{equation}
  \label{eq:H0}
  H_0^{\mathrm{3D}} = \int \! d\vec{r} \left\{ 
    \hat{\psi}^{\dagger}(\vec{r}) h_1
    \hat{\psi}(\vec{r}) 
    + \frac{g}{2}
    \hat{\psi}^{\dagger}(\vec{r})
    \hat{\psi}^{\dagger}(\vec{r}) 
    \hat{\psi}(\vec{r}) \hat{\psi}(\vec{r}) \right\}.
\end{equation}
The atomic field operators $\hat\psi(\vec r)$ and
$\hat\psi^{\dagger}(\vec r)$ annihilate and create an atom at
position $\vec r$ and they satisfy the equal time commutation relations
\begin{equation}
  \label{eq:commu}
  \left[ \hat\psi(\vec r),\hat\psi^{\dagger}(\vec r')\right]
  = \delta^3(\vec r - \vec r')
\end{equation}
The single particle Hamiltonian $h_1$ is the one appropriate for a
single atom of mass $m$ in the external trapping potential
$V_{ext}$
\begin{equation}
  \label{eq:h_1}
  h_1 = - \frac{\hbar^2}{2m}\vec{\nabla}^2+V_{ext}(\vec{r}).
\end{equation}
The $g$ term in Eq.(\ref{eq:H0}) describes the
collisional two body interaction of the atoms. The simple contact form
is an approximation appropriate for cold, dilute
gases~\cite{huang87:_statis_mechan}. The value of the strength
parameter is $g=4\pi \hbar a_s / m$ where $a_s$ is the s-wave
scattering length.  

In most experiments the trapping potential can be approximated well by
a three dimensional harmonic oscillator characterized by three
frequencies $\omega_x$, $\omega_y$ and $\omega_z$. If one of these
frequencies is significantly smaller than the other two the atoms will
form an elongated cloud and we describe such an effectively one
dimensional system by a single frequency
$\omega$~\cite{steel98:_dynam_bose_einst}. 

We shall work with dimensionless equations and choose to
measure time in
  units of $\omega^{-1}$, lengths in units of $a_0=\sqrt{\hbar/\omega
    m}$, and energy in units of $\hbar \omega$. We are then left with
\begin{equation}
  \label{eq:H0_1D}
  H_0 = \int \! dx \left\{ \hat{\psi}^{\dagger}(x) h_1 \hat{\psi}(x) 
    + \frac{g}{2} \hat{\psi}^{\dagger}(x) \hat{\psi}^{\dagger}(x) 
    \hat{\psi}(x) \hat{\psi}(x) \right\}
\end{equation}
where
\begin{equation}
  \label{eq:h_1_1D}
  h_1 = - \frac{1}{2}\frac{\partial^2}{\partial x^2}+\frac{1}{2}x^2.
\end{equation}
and where $g=4 \pi a_s/a_0$ is a dimensionally correct, one
dimensional interaction
strength~\cite{g_1d}.
 
\subsection{Photodissociation from a molecular condensate}
\label{sec:pd}
Starting from a molecular condensate (prepared by photo-association or
by other means) we can imagine to photo-dissociate or ``down convert''
the molecules to pairs of free atoms. A suitable Hamiltonian to
describe photo-dissociation is
\begin{equation}
  \label{eq:H_PD}
  H_{PD} = \frac{1}{2} \int \! dx \; \left\{
  b(x,x',t) \hat\psi^{\dagger}(x)\hat\psi^{\dagger}(x') 
  + b^*(x,x',t) \hat\psi(x) \hat\psi(x')
  \right\}.
\end{equation}
This Hamiltonian clearly creates and annihilates atoms in pairs.  A
full description should also include the molecules such that each
atomic pair creation would be accompanied by the annihilation of a
molecule. Here we will assume the molecular condensate to be large and
the number of molecules removed to be small. In that case it is
reasonable to describe the molecular condensate by a time-independent
c-number field. The function $b$ in Eq.(\ref{eq:H_PD}) is the product
of this field and of the coupling to a position dependent laser field
which is also assumed to be classical. To represent the position
dependence of the molecular condensate and of the laser field, the
relative wavefunction of the pair of atoms, and the time dependence of
the laser field, we use the ansatz
\begin{equation}
  \label{eq:b_def}
b(x,x',t)=\frac{ B  }{ 2 \pi \sigma_r \sigma_{cm} }
  \exp \left( - 2i \Delta t \right) 
  \exp \left( -\frac{1}{2} \frac{(x - x')^2}{\sigma_r^2} \right)
\exp \left( -\frac{1}{2} \frac{(\frac{1}{2} ( x + x'))^2
  }{\sigma_{cm}^2} \right)
\end{equation}
where $\sigma_r \ll \sigma_{cm} \cong a_0$.

\subsection{Operator equations of motion}
\label{sec:eqn_evol}

A successful tool for the description of a Bose condensate is the
Gross-Pitaevskii equation (GPE). It can be obtained from the Heisenberg
equation of motion for the atomic field $\hat\psi(x,t)$ by taking
average values and by replacing the mean value of an operator product
by the product of mean values. In our case the Heisenberg
equation reads:
\begin{equation}
  \label{eq:psi}
  i \frac{\partial \hat\psi(x,t)}{\partial t}
  = \left( -\frac{1}{2}\frac{\partial}{\partial x} 
    + \frac{1}{2} x^2 + g \hat\psi^\dagger(x,t) \hat\psi(x,t)
  \right) \hat\psi(x,t) + \int dx' \; b(x,x',t) \hat\psi^\dagger(x',t).
\end{equation}
The last term in this equation is due to the exchange of atom pairs
with the molecular condensate {\it via} photo-dissociation. 

It is
 easy to see that using the average of Eq.(\ref{eq:psi}) has
 shortcomings when we try to describe 
photo-dissociation: If we start from the atomic vacuum defined by
\begin{eqnarray}
  \label{eq:vaccum}
  \hat\psi(x,0) |0\rangle &=& 0 
\end{eqnarray}
the incoupling term on the right hand side has vanishing average
value. Therefore $\langle \hat\psi(x,t) \rangle$ will stay zero also at
all later times and no useful information can be extracted.

To gain knowledge of the state created we therefore proceed
to study expressions quadratic in the field operators. To shorten
notation we define
\begin{eqnarray}
  \label{eq:def_rhat}
  \hat R(x,y,t) &\equiv& \hat\psi^\dagger(x,t)\hat\psi(y,t) \\ 
  \label{eq:def_shat}
  \hat S(x,y,t) &\equiv& \hat\psi(x,t)\hat\psi(y,t) .
\end{eqnarray}
For these operators we get the following Heisenberg equations of motion:
\begin{eqnarray}
  \label{eq:rhat_eom}
  i\frac{\partial \hat R(x,y,t)}{\partial t} &=&
  \left( \frac{1}{2} \frac{\partial^2}{\partial x^2}
    -\frac{1}{2} x^2
    -\frac{1}{2} \frac{\partial^2}{\partial y^2}
    +\frac{1}{2} y^2
  \right) \hat R(x,y,t)
  \nonumber \\
  && + g\left( \hat R(y,y,t) 
    - \hat R(x,x,t)  \right) \hat R(x,y,t)
  \nonumber \\
  && + \int dz \; \left\{
  b(z,y,t) \hat S^\dagger(x,z,t)
  -b^*(x,z,t) \hat S(z,y,t)
  \right\}
\end{eqnarray}
and
\begin{eqnarray}
  \label{eq:shat_eom}
  i\frac{\partial \hat S(x,y,t)}{\partial t} &=&
   \left( -\frac{1}{2} \frac{\partial^2}{\partial x^2}
    +\frac{1}{2} x^2
    -\frac{1}{2} \frac{\partial^2}{\partial y^2}
    +\frac{1}{2} y^2
  \right) \hat S(x,y,t)
  \nonumber \\
  && + g \left( \delta(x-y) + \hat R(x,x,t)
    +\hat R(y,y,t)
  \right) \hat S(x,y,t)
  \nonumber \\
  && + \int dz \; \left\{
  b(z,y,t) \hat R(z,x)
  +b(x,z,t) \hat R(z,y)
  \right\}
  +b(x,y,t).
\end{eqnarray}
Note that $b(x,y,t)$ now appears as an inhomogeneous source term in the $\hat S$
equation. This guarantees a nontrivial behaviour when we take
averages. Note also that when taking averages we have a problem with
the interaction terms which will couple the second order expectations
to fourth order expectations. These fourth order terms have to be
factorized in some approximate way to obtain a closed set of equations.

\subsection{c-number equations}
\label{sec:cnum_eq}

\subsubsection{Exact equations for $g=0$}
\label{sec:g_0}
When $g=0$, Eq.(\ref{eq:rhat_eom}) and Eq.(\ref{eq:shat_eom}) can be
reduced to two coupled linear equations for the moments
\begin{equation}
  \label{eq:r_def}
R(x,y,t) \equiv \langle \hat R(x,y,t) \rangle
\mathrm{~~,~~}
S(x,y,t) \equiv \langle \hat S(x,y,t) \rangle.    
\end{equation}
They read:
\begin{eqnarray}
  \label{eq:r_eom}
i\frac{\partial R(x,y,t)}{\partial t}&=&
 \left( \frac{1}{2} \frac{\partial^2}{\partial x^2}
    -\frac{1}{2} x^2
    -\frac{1}{2} \frac{\partial^2}{\partial y^2}
    +\frac{1}{2} y^2
  \right) R(x,y,t)
  \nonumber \\
  && + \int dz \; \left\{
    b(z,y,t) S^*(x,z,t)
    -b^*(x,z,t) S(z,y,t)
  \right\}\\
  \label{eq:s_eom}
i\frac{\partial S(x,y,t)}{\partial t}&=&
\left(- \frac{1}{2} \frac{\partial^2}{\partial x^2}
    +\frac{1}{2} x^2
    -\frac{1}{2} \frac{\partial^2}{\partial y^2}
    +\frac{1}{2} y^2
  \right) S(x,y,t)
  \nonumber \\
   && + \int dz \; \left\{
  b(z,y,t) R(z,x)
  +b(x,z,t) R(z,y)
  \right\}
  \nonumber \\ 
  &&+b(x,y,t). 
\end{eqnarray}
Moreover, $R$ and $S$ uniquely determine all higher order expectation
values. This can be seen in a number of ways. One is to note that the
Wigner distribution is a multi-dimensional gaussian distribution fully
characterized by its second order
moments~\cite{kitamura99:_squeez,gardiner91:_quant_noise}.  Another
follows from the observation that when $g=0$, the Heisenberg equation
of motion (\ref{eq:psi}) implies that $\hat\psi(x,t)$ can at all times
be expressed as a linear combination of the initial values
\begin{eqnarray}
  \label{eq:psi_p_psi_m}
  \hat\psi(x,t) &=&\int dy \; \left\{
    f(x,y,t)\hat\psi(y,0)+g(x,y,t)\hat\psi^\dagger(y,0)
  \right\} .
\end{eqnarray}
Eq.(\ref{eq:psi_p_psi_m}), its hermitian conjugate and the fact that
our system starts in the vacuum state then suggest the following scheme for
calculation of any operator product at arbritrary $t$: Use the
commutation relation (\ref{eq:commu}) to move all $\hat\psi(x,0)$ to
the right of any $\hat\psi^\dagger(y,0)$ (normal ordering). Of all the
terms produced in this process only the ones consisting entirely of
c-numbers are nonzero as the vacuum expectation of any normal ordered
product of operators vanishes in the vacuum state. To evalute the
c-number terms we formally need to calculate integrals of products of
the $f$ and $g$ functions of Eq.(\ref{eq:psi_p_psi_m}). It is,
however, not difficult to see that these integrals factorize and that
the factors are exactly the ones involved in calculating expectation
values of products of only two field operators. The end result is that
the average of any operator product is replaced by a sum of all
possible factorizations into two-operator
expectations 
\begin{eqnarray}
  \label{eq:wick_ex}
  \langle \hat\psi^\dagger(x_1) \hat\psi^\dagger(x_2) \hat\psi(x_3) \hat\psi(x_4) \rangle
  &=& 
  \langle \hat\psi^\dagger(x_1) \hat\psi^\dagger(x_2) \rangle
  \langle \hat\psi(x_3) \hat\psi(x_4)  \rangle
  \nonumber \\ 
  && + \langle \hat\psi^\dagger(x_1) \hat\psi(x_3) \rangle 
  \langle \hat\psi^\dagger(x_2) \hat\psi(x_4) \rangle
  \nonumber \\
  && + \langle \hat\psi^\dagger(x_1) \hat\psi(x_4) \rangle 
  \langle \hat\psi^\dagger(x_2) \hat\psi(x_3) \rangle.
\end{eqnarray}
This is a simple version of Wick's
theorem~\cite{huang98:_quant_field_theor}.

\subsubsection{Approximate equations for $g\not = 0$ }
\label{sec:app_eq_gnotzero}

When $g\not=0$ we have to include the interaction term of
Eqs.(\ref{eq:rhat_eom},\ref{eq:shat_eom}) in
Eqs.(\ref{eq:r_eom},\ref{eq:s_eom}). Unfortunately, the decomposition
Eq.(\ref{eq:psi_p_psi_m}) is no longer exact, and there is no simple
way to reduce the mean values of four-operator products to products of
$R$ and $S$.  Rather than the simple replacement, {\it e.g.}, $\hat
R(y,y) \hat R(x,y) \rightarrow R(y,y)R(x,y)$, we choose to apply
the Wick prescription as this is correct to lowest order.  We then get
\begin{eqnarray}
  \label{eq:r_gterms}
   g\left\langle \hat R(y,y,t) \hat R(x,y,t)
      - \hat R(x,x,t) \hat R(x,y,t) \right\rangle
  &  \rightarrow & 
   g \left( S^*(x,y,t)S(y,y,t) -S(x,y,t)S^*(x,x,t) \right)
  \nonumber \\
  && +2g \left(R(y,y,t) - R(x,x,t)\right) R(x,y,t) 
  \\
  \label{eq:s_gterms}
     g \left\langle \left( 
        \hat R(x,x,t) +\hat R(y,y,t) +\delta(x-y) 
      \right) \hat S(x,y,t) \right\rangle
   & \rightarrow &
  2g \left(R(y,y,t) + R(x,x,t)\right) S(x,y,t) 
  \nonumber \\
  && + g \left( R(x,y,t)S(x,x,t) +R^*(x,y,t)S(y,y,t) \right)
  \nonumber \\
  && + g \delta(x-y) S(x,y,t)
\end{eqnarray}

When these expressions are inserted into Eqs.(\ref{eq:r_eom},\ref{eq:s_eom}),
we arrive at the equations we want to solve numerically. We use a
split-step approach where the kinetic energy is treated by a Fourier
method. The remaining terms are
dealt with by a fourth order Runge-Kutta scheme. In this
one-dimensional problem the equations are quite manageable.

\subsection{The positive $P$ pseudo-probability distribution}
\label{sec:Pplus}

Pseudo-probablity distributions (PPD's) are well-established tools in
quantum mechanics. The most well known of the distributions are the
Glauber-Sudarshan P-function and the Wigner function but especially in
quantum optics a number of other distributions have also been useful.
Common to all the PPD's is that they provide the expectation values of
properly ordered operator products as weighted c-number averages. The
\emph{positive $P$
  distribution}~\cite{gardiner91:_quant_noise,steel98:_dynam_bose_einst,drummond99:_quant_bose_einst} 
($P_+$) that we will be using here gives the expection values of
normally ordered products by replacing $\hat\psi$ by a c-number
function $\psi_1$ and $\hat\psi^\dagger$ by a c-number function
$\psi_2$, {\it e.g.}:
\begin{equation}
  \label{eq:Pp_ex}
  \langle \hat{\psi}^{\dagger}(x,t)\hat{\psi}(y,t) \rangle = 
  \int d [ \psi_1 ] d [ \psi_2 ] \psi_2(x)\psi_1(y) P_+ [ \psi_1,\psi_2,t ].  
\end{equation}
Note that $P_+$ is the joint distribution of two spatial functions and
it is therefore an immensely complicated functional in general. It
satisfies a multi-dimensional Fokker-Planck equation which, however,
opens the door to a Monte Carlo sampling as we can translate the
Fokker-Planck equation for the distribution to Langevin equations for
stochastic realizations of $\psi_1(x,t)$ and $\psi_2(x,t)$. These
equations resemble the GPE but they are coupled and they contain noise
terms. A derivation of the equations without incoupling is given
in~\cite{steel98:_dynam_bose_einst} (see
also~\cite{carusotto00:_stoch_wave}) and the inclusion of incoupling
is straightforward. In the notation of stochastic differential
equations the equations can be written
\begin{eqnarray}
    i
    d \psi_1(x) &=&  
  \left( h_1 \psi_1(x)
    +g\psi_2(x)\psi_1(x)\psi_1(x)
  \right) dt \\
  && +\int b(x,x')\psi_2(x') dx'
  + dW_1(x)  
  \label{eq:langevin1} \\
  -i
    d \psi_2(x) &=& 
  \left( h_1 \psi_2(x)
    +g\psi_1(x)\psi_2(x)\psi_2(x)
  \right) dt \\
   && + \int  b^*(x,x') \psi_1(x') dx'
   +  dW_2(x)
\label{eq:langevin2} 
\end{eqnarray}
where $h_1$ is still defined in Eq.(\ref{eq:h_1}) and the noise terms
are gaussian and given by
\begin{eqnarray}
  \label{eq:wiener}
  \langle dW_{1,2}(x,t) \rangle &=& 0 \\
  \label{eq:wiener12}
  \langle dW_{1}(x,t) dW_{2}(x',t') \rangle &=& 0 \\
  \label{eq:wiener11}
  \langle dW_{1}(x,t) dW_{1}(x',t') \rangle &=& i dt
  \left( b(x,x',t) + g \psi_1(x,t) \delta(x-x')\right)\delta(t-t')
   \\
   \label{eq:wiener22}
  \langle dW_{2}(x,t) dW_{2}(x',t') \rangle &=& -i dt
  \left( b^*(x,x',t) + g \psi_1(x,t)\delta(x-x') \right)\delta(t-t')
   . 
\end{eqnarray} 
By numerically simulating
Eqs.(\ref{eq:langevin1},\ref{eq:langevin2}) we are able to calculate
expectation values of arbitrary, normally ordered field operator
products with the only approximation that the results are
subject to sampling errors due to the use of finite
ensembles. We describe in the appendix our procedure to synthesize the
noise $dW$ in our simulations.

The crucial drawback of the method is the well know sudden divergence
in some of the unphysical moments of the $P_+$
distribution\cite{carusotto00:_stoch_wave,posp_diverge}.
Unphysical moments exist because the translation from operator
products to products of $\psi_1$ and $\psi_2$ never involves
$\psi_1^*$ and $\psi_2^*$. This leaves some room for $P_+$ to behave
badly and unfortunately it exploits this freedom.  In the wavefunction
realizations, some wavefunctions diverge or they make very large
excursions that are difficult to follow numerically and which make a
devastating impact on the sampling error.  The $P_+$ Monte Carlo
method is therefore limited to short times where only few atoms have
been created and nonlinear effects are still small. This suits our
purpose, since we are only interested in short time dynamics, and we
shall trust the result produced by the Langevin equations as long as
none of the wavefunctions in the ensemble have escaped the region
where we have confidence in our integration algorithm.

\section{Results}
\label{sec:results}

In this section we show results for some of the quantities of interest
that we are able to calculate in our model. Although the main new
feature lie in the quantum correlations we first show
a very classical quantity, namely the density profile. We then proceed
to look at the eigenvalues of the one-particle density matrix. The
largest of these eigenvalues defines the condensate fraction and the corresponding
eigenvector is the condensate wavefunction. Finally we turn to a
two-body quantity, the second order correlation function.   

\subsection{The density profile and the number of atoms}
\label{sec:density}

\begin{figure}[htbp]
  \begin{center}
   \resizebox{8.5cm}{!}
   {
   \includegraphics{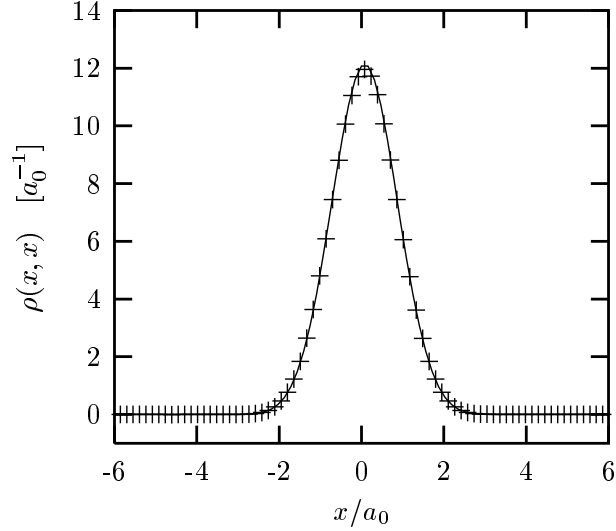}
   }
  \end{center}
  \caption{The density profile at time $\omega t=2.4$ for interaction strenght
    $g=0.01$ and incoupling parameters $B=3.0\hbar^2/m$,
    $\Delta=\omega$, $\sigma_{cm}=a_0$ and $\sigma_r=0.2a_0$. The
    solid curve shows the results of the coupled R\&S equations, the
    crosses are obtained with the $P_+$ simulations.}
  \label{fig:density}
\end{figure}

The atomic density is given by the diagonal elements of the one-body
density operator in the position representation, that is
\begin{equation}
  \label{eq:density}
  \rho(x,x) =  \langle \hat\psi^{\dagger}(x) \hat\psi(x) \rangle 
  = R(x,x) = \overline{\psi_2(x)\psi_1(x)},
\end{equation}
where the overbar in the last expression denotes the average over many
realizations of the stochastic $\psi_1(x,t)$ and $\psi_2(x,t)$.  In
Fig.~\ref{fig:density} is shown a typical plot of this profile at
$\omega t=2.4$. It has the characteristic gaussian shape of the
harmonic oscillator ground state and as we shall see in
Sec.~\ref{sec:cond_frac} a large fraction of the atoms indeed occupy a
common wavefunction close to this state. The R\&S
equations~(\ref{eq:r_eom},\ref{eq:s_eom}) with the interaction
terms~(\ref{eq:r_gterms},\ref{eq:s_gterms}) give results in excellent
agreement with the $P_+$ simulations.

\begin{figure}[htbp]
  \begin{center}
   \resizebox{8.5cm}{!}
   {
   \includegraphics{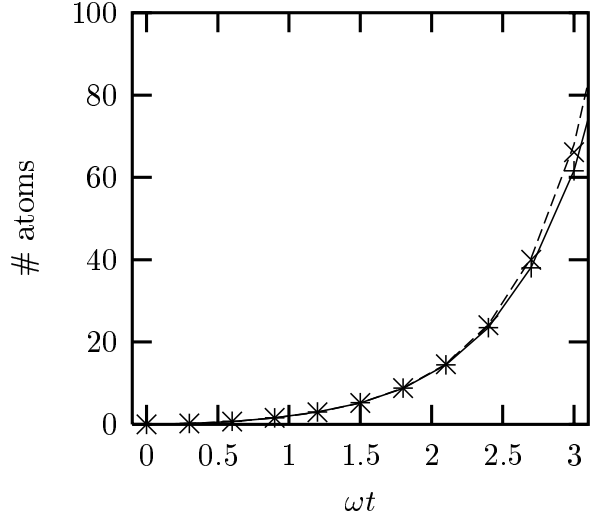}
   }
  \end{center}
  \caption{The number of photo-dissociated atoms as a function of
    time. The solid and dashed curves show the results of the R\&S
    equations for $g=0.00$ and $g=0.01$ while the symbols $+$ and
    $\times$ indicate the corresponding results of the $P_+$ Langevin
    equations. Incoupling as in Fig.\ref{fig:density}.}
  \label{fig:natoms}
\end{figure}

The total number of photo-dissociated atoms is obtained as the trace
of the one-body density-operator or, according to
Eq.(\ref{eq:density}), simply as
\begin{equation}
  \label{eq:N_from_R}
  N = \int dx \; \rho(x,x).
\end{equation}
In Fig.~\ref{fig:natoms} this number is shown as a function of time
for $g=0$ and for $g=0.01$. The agreement between the R\&S equations
and the $P_+$ method is seen to be quite good.


\subsection{The condensate fraction and wavefunction}
\label{sec:cond_frac}
By diagonalizing the one-body density matrix we obtain an orthonormal
basis of single-particle states. The eigenvalues correspond to the
populations of these states and in the case of a condensate one of
these eigenvalues dominates, {\it i.e.}, most particles occupy the
same state. A dynamical picture of the condensation process is the
Bose-enhancement of the scattering into the most occupied state. Here
we expect a similar effect to take place. At first several states of
the system are occupied by the atoms created. As the number of atoms
grows the stimulated character of the creation becomes more important
and a mode competition results in one mode being preferentially
occupied.

\begin{figure}[htbp]
  \begin{center}
   \resizebox{8.5cm}{!}
   {
   \includegraphics{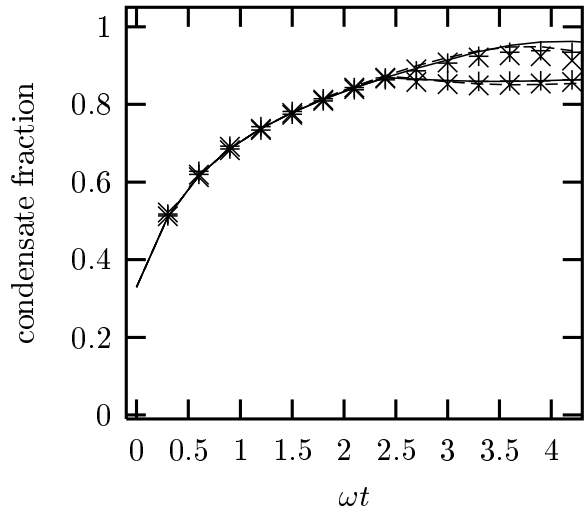}
   }
  \end{center}
  \caption{The condensate fraction, {\it i.e.}, the ratio of the largest
    eigenvalue of the one-body density matrix to the total number of
    atoms (trace of the one-body density matrix). Solid and dashed
    curves shown the results of the R\&S equations
    for $g=0.01$ and for $g=0.02$. The corresponding results of the
    $P_+$ simulations are indicated by the symbols $+$ and
    $\times$. The lower sets of data with the same symbols show the
    results when the incoupling is stopped at 
    $\omega t = 2.4$.}
  \label{fig:condfrac}
\end{figure}

In Fig.\ref{fig:condfrac} we show the condensate fraction, {\it
i.e.}~the ratio of the largest eigenvalue of the one-body density
matrix to the sum of the eigenvalues for different values of the
interaction strenght $g$. It is seen that as expected the condensate
fraction is in general an increasing function of time. The effects of
interactions are rather small at these low atom numbers. Note that
unlike studies of stationary condensates at $T=0$, where interactions
are responsible for the breakdown of a simple product state ansatz for
the system and the existence of atoms outside the condensate, our
incoupling by itself produces atoms both in the condensate and outside
the condensate. In fact, our calculations show that the second-largest
eigenvalue accounts for most of the atoms which are not in the
condensate.

\begin{figure}[htbp]
  \begin{center}
   \resizebox{8.5cm}{!}
   {
   \includegraphics{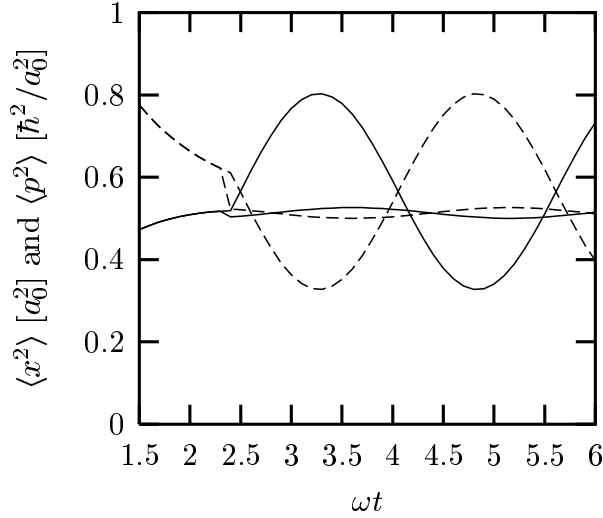}
   }
  \end{center}
  \caption{The freely evolving expectation values of $\langle x^2
    \rangle$ (solid curve) and $\langle p^2 \rangle$ (dashed curve)
    are compared to the results obtained after
    application of a $\delta$-kick to match the wavefunction with the
    ground state in the trap. If we apply a properly chosen kick just
    as we stop the incoupling at $\omega t = 2.4$ the oscillations can
    be almost completely removed.}
  \label{fig:x2p2}
\end{figure}

As for the condensate wavefunction we see an interesting phenomenon:
Although the density profile associated with the condensed part of the
one-body density matrix is close to that of the trap ground state, the
condensate wavefunction is in fact not stationary. The atoms have
condensed into a state more resembling a squeezed state\footnote{This
position-momentum squeezing should not be confused with the the
atom-field squeezing discussed later.} and if the incoupling is
stopped the wavefunction widths show an oscillating behavior. In Fig.
\ref{fig:x2p2} we show $\langle \hat x^2 \rangle$ and $\langle \hat
p^2 \rangle$ of the condensate wavefunction as a function of time. We
see that at $\omega t = 2.4 $ when the incoupling is stopped, the
wavefunction is too wide in momentum space as compared to the ground
state of the trap ($\langle \hat p^2 \rangle > 1/2$).

One way to avoid this oscillation is to apply $\delta$-kick
cooling~\cite{delta_kick_cooling} to the
system. This procedure is efficient if there is a linear correlation
between position and momentum. In the original suggestion the
correlation between position and momentum is brought about by free
expansion, but an examination of the condensate wavefunction shows
that we have a similar correlation here. The idea is to apply a tight,
harmonic trapping potential for a short time interval. If this
interval is so short that any changes in position can be ignored, the
effect is simply a momentum kick also varying linearly with position.
Ideally, this kick brings all the particles to rest. In
Fig.~\ref{fig:x2p2} we demonstrate that the procedure is effective in
our problem. The results are shown for $g=0$, but a similar reduction
is achieved for non-vanishing $g$.

\subsection{The second order correlation function
  \protect{$g^{(2)}(x,y)$}}
\label{sec:g2}

More detailed information about the quantum state of the system is
desired and available, and a natural quantity to consider is the
second order correlation function
\begin{equation}
  \label{eq:g2_def}
    g^{(2)}(x,y) \equiv \frac{ \langle
  \hat\psi^\dagger(x)\hat\psi^\dagger(y)\hat\psi(y)\hat\psi(x) \rangle
  } { \langle \hat\psi^\dagger(x)\hat\psi(x) \rangle \langle
  \hat\psi^\dagger(y)\hat\psi(y) \rangle }.
\end{equation}
It measures the probability to find two atoms at positions $x$ and $y$
normalized to the single-particle densities at the two positions. For
a thermal state $g^{(2)}=2$ while for a coherent state $g^{(2)}=1$.

As $g^{(2)}$ involves the expectation value of a product of four field
operators we are faced with similar factorization problem as when we
derived the R\&S equations. Again we will resort to the Wick
prescription although it should be realized that this is only exact
for states obtained with $g=0$.

We have already evaluated the expectation in the enumerator of
Eq.(\ref{eq:g2_def}) in terms of $R$ and $S$ in the gaussian case.
This was done in Eq.(\ref{eq:wick_ex}) and we get for the second order
correlation function
\begin{equation}
  \label{eq:g2_wick}
  g^{(2)}(x,y) = 1+\frac{\left| R(x,y) \right|^2+\left| S(x,y)
  \right|^2}{R(x,x)R(y,y)}.
\end{equation}

In contrast with the R\&S equations the $P_+$ method has no problems
handling expectation values like the enumerator of
Eq.(\ref{eq:g2_def}), and $g^{(2)}$ can be determined exactly up to
sampling errors. At relatively short times and low atom numbers we
have therefore an excellent tool to obtain exact results even for
$g\not=0$.

\begin{figure}[htbp]
  \begin{center}
    \resizebox{8.5cm}{!}
    {
    \includegraphics{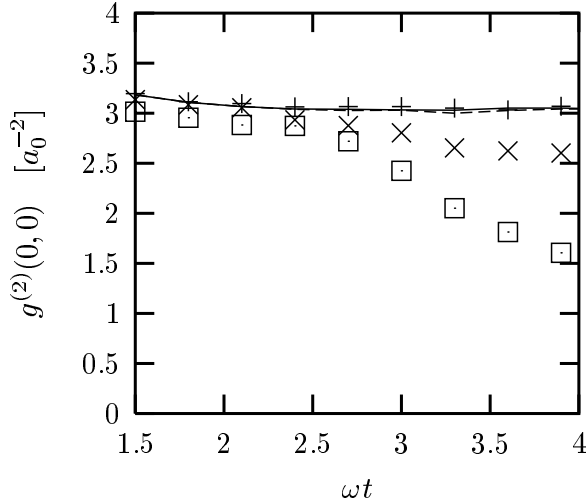}
   }
  \end{center}
  \caption{The second order correlation function $g^{(2)}(x,y)$
    evaluated at $y=x=0$ as a function of time. The solid and dashed
    lines show the results of the R\&S equations for $g=0$ (exact) and
    $g=0.02$. The symbols $+$ and $\times$ show the results of the
    $P_+$ simulations with $g=0.00$ and $g=0.02$. The photo-dissociation
    stops at $\omega t = 2.4$. For $g \not= 0$,  here is a strong discrepancy between
    the results of the R\&S equations and the $P_+$ at times beyond
    $\omega t = 2.5$.  The effect is clearly dependent on the
    interaction. The symbol $\boxdot$ show $P_+$ results obtained when
    $g=0.02$, and when the photodissociation is not interrupted.}
  \label{fig:g2}
\end{figure}

In Fig.~\ref{fig:g2} we show a plot of $g^{(2)}(0,0)$ as a function of
time for various values of $g$. The central value slightly above 3
indicates a strong bunching effect where two atoms are more likely to
be found close together than in a coherent or a thermal state. This
result can be compared with the analytical expression for a single
mode squeezed state, generated by the Hamiltonian
$\beta\left(\hat{a}^2+{\hat{a}^\dagger\mbox{}}^2\right)$:
\begin{equation}
  \label{eq:single_mode_squeezed}
  g^{(2)}= \frac{\langle \hat a^\dagger \hat a^\dagger \hat a \hat a
  \rangle} {\langle \hat a^\dagger \hat a \rangle^2} = 3 +
  \frac{1}{\langle \hat a^\dagger \hat a \rangle}
\end{equation}

In the figure we plot both results of the R\&S equations using
Eq.(\ref{eq:g2_wick}) and the exact $P_+$ results. Good agreement is
found between the two approaches until $\omega t \cong 2.5$ and
hereafter the R\&S equations fail to capture a decrease in the value
of $g^{(2)}$. This decrease indicates a threshold effect that we will
discus in the next section.

\subsection{Threshold effect}
\label{sec:threshold}

In the semi-classical treatments of the laser and of the
parametric oscillator, one identifies a threshold 
in the stationary balance between gain and loss; 
the fields shift from fluctuations around zero to fluctuations
around finite intensities \cite{threshold}. Above threshold
these optical systems have smaller relative fluctuations of the
intensity, and it is natural to expect a similar threshold behaviour
in our model. There is a seemingly important difference between
our model and the optical systems in the fact that we do
not have an explicit dissipative mechanism. It has been
known for a long time, and it has been demonstrated explicitly
for a large number of physical systems, however, that quite 
generically, the interactions in many-body systems lead to 
ergodicity of eigenstates and a dynamical relaxation without
coupling to an external bath. For a recent review, see \cite{guhr}.

\begin{figure}[htbp]
  \begin{center} 
    \resizebox{8.5cm}{!}
    {
      \includegraphics{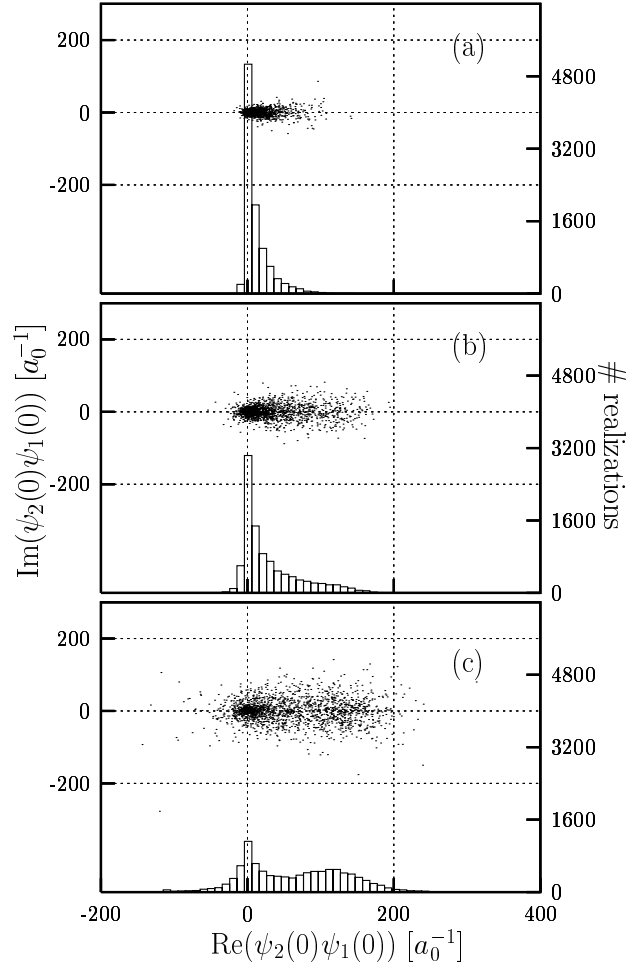}
      }
  \end{center} 
  \caption{Scatter plots of $\psi_2(0)\psi_1(0)$ from the $P_+$
    realizations for $\omega t = 2.4$ (a), $\omega t = 3.0$ (b), and
    $\omega t =3.6$ (c).  The average of this quantity is $\langle
    \hat\psi^\dagger(0)\hat\psi(0) \rangle$, the central density in
    the trap and higher order, normally ordered, moments $\langle
    \hat{\psi}^\dagger\mbox{}^n \hat{\psi}^n \rangle$ are similar
    averages $\overline{\left(\psi_2(0)\psi_1(0)\right)^n}$. For
    clarity, we show in the bottom of the plots histograms of
    $\mathrm{Re}(\psi_2(0)\psi_1(0))$. It is seen how the character of
    the state changes with time as a second maximum in the
    distribution develops at a value different from zero.}
  \label{fig:complexN}
\end{figure}

The R\&S equations with their underlying assumption of a gaussian
Wigner distribution, centered around vanishing atomic field, are
clearly unable to describe correctly the system around and above
threshold, but the $P_+$ simulations are exact, and the discrepancy
between the two methods is thus most likely explained by a threshold
effect.  To investigate more closely the threshold hypothesis, we show
in Fig.~\ref{fig:complexN} scatter plots of $\psi_2(0)\psi_1(0)$ at
$\omega t = 2.4$, $3.0$ and $3.6$ for a situation with $g=0.02$.
From photodetection theory we can deduce 
the following expression for the atom number distribution
\begin{equation}
  \label{eq:Pn}
  P_n(t)=\overline{\frac{[\int \psi_2(x,t) \psi_1(x,t)dx]^n}{n!}
    \exp\left(-\int \psi_2(x,t)\psi_1(x,t)dx\right)}.
\end{equation}
This expression, however, exhausts the statistical precision of the
$P_+$ method, since it involves higher moments of the simulated
amplitudes, which yield larger and larger fluctuations.  Instead, we
heuristically present histograms in the figures of how the real parts,
$\mathrm{Re}(\psi_2(0)\psi_1(0))$ are distributed, and these
histograms are in good qualitative agreement with our picture of a
bifurcation of the solution when we reach threshold in the process.
It is seen how the distribution at $\omega t = 2.4$ is strongly peaked
at zero with an exponential tail along the the real axis. At $\omega t
= 3.0$ this tail extends to larger values, and a shoulder around 120
atoms/$a_0$ starts to appear, and at $\omega t = 3.6$ a second maximum
has developed. This explains the lowering of $g^{(2)}$ seen in
Fig.~\ref{fig:g2}.

\section{Application of a squeezed condensate}
\label{sec:appli}

In this section we analyse a possible application of the state created
in our model. We show how its peculiar statistical properties can be
utilized to produce precise measurements. Our suggestion is in direct
analogy with the use of a \emph{squeezed vacuum} in quantum optics
experiments with beam-splitters. When a beam in such an experiment is
incident on a 50/50 beam-splitter and is split in two, the analysis of
the noise properties (quantum fluctuations) of the two daughter beams
depends crucially on realizing that the incoming beam is not only
split at the beam-splitter, it is actually mixed with vacuum coming in
from the back-side of the mirror. By replacing this vacuum by a
squeezed state we may control the statistical properties of the
daughter beams.

The matter wave analogue of the beam-splitter could in our case be a
laser pulse which is able to coherently change the internal state of
atoms. Such a pulse can drive each atom into a
superposition of two internal states.

Suppose now that we have two internal states of the
atoms, state $a$ and state $b$, and that we apply a $\pi/2$-pulse. We
then have in the Heisenberg picture
\begin{eqnarray}
  \label{eq:pi2}
  \hat\psi_a & \rightarrow & \hat\psi'_a = 
   \frac{1}{\sqrt{2}} \left( \hat\psi_a + \hat\psi_b \right)\\
  \hat\psi_a & \rightarrow & \hat\psi'_b =
  \frac{1}{\sqrt{2}} \left( \hat\psi_a - \hat\psi_b \right).
\end{eqnarray}
The total number operators of atoms in state $a$ and state $b$ after the
pulse are thus given by
\begin{eqnarray}
  \label{eq:Na_hat}
  \hat N'_a &=& \int dx \hat{\psi'}_a^\dagger(x) \hat\psi'_a(x)
  \nonumber \\ 
  &=& \frac{1}{2}\hat N_a + \frac{1}{2} \hat N_b
  + \frac{1}{2}\int dx \left\{ \hat\psi_a^\dagger(x) \hat\psi_b(x)
      +\hat\psi_b^\dagger(x) \hat\psi_a(x) \right\} \\
  \label{eq:Nb_hat}
  \hat N'_b &=& \int dx \hat{\psi'}_b^\dagger(x) \hat\psi'_b(x)
  \nonumber \\ 
  &=& \frac{1}{2}\hat N_a + \frac{1}{2} \hat N_b
  - \frac{1}{2}\int dx \left\{ \hat\psi_a^\dagger(x) \hat\psi_b(x)
      +\hat\psi_b^\dagger(x) \hat\psi_a(x) \right\}.
\end{eqnarray}  
We will now concentrate on the \emph{difference} in the
number of atoms in the two states,
\begin{eqnarray}
  \label{eq:Na-Nb_hat}
  \hat N'_a - \hat N'_b &=&
  \int dx \left\{ \hat\psi_a^\dagger(x) \hat\psi_b(x)
    + \hat\psi_b^\dagger(x) \hat\psi_a(x) \right\}.
\end{eqnarray}
We will also let $\hat\psi_a$ initially describe a large condensate
while $\hat\psi_b$ describes our photodissociated state. That is, we
assume that the photodissociation is to the internal state $b$, while
at the time of the $\pi/2$-pulse these atoms are overlapped with a
large normal condensate in the internal state $a$. The large
condensate is assumed to be in a coherent state with a definite phase
, {\it e.g.}
\begin{eqnarray}
  \label{eq:coh_def}
  \hat\psi_a(x) |\rangle & = & e^{i\theta}\phi_a(x) |\rangle \\  
  \langle | \hat\psi_a^\dagger(x) & = & e^{-i\theta}\phi_a(x) \langle | 
\end{eqnarray}
with $\phi_a$ real. The mean number of atoms is given by
\begin{equation}
  \label{eq:Na}
  N_a=\langle \hat N_a \rangle = \int dx \; \phi_a^2(x).
\end{equation}

Using Eq.(\ref{eq:Na-Nb_hat}) we find
\begin{equation}
  \label{eq:Na-Nb}
  \left\langle \hat N'_a - \hat N'_b \right\rangle = 0
\end{equation}
and
\begin{equation}
  \label{eq:Na_Nb_2}
  \left\langle \left( \hat N'_a - \hat N'_b \right)^2 \right\rangle =  
  N_a + N_b + \int \!\! \int dxdy \; 
  \phi_a(x) \phi_a(y)
  \mathrm{Re}\left[  R_b(x,y,) + e^{-2i\theta} S_b(x,y,) \right] 
  .
\end{equation}
Ordinary vacuum in the $b$-state is the special case with
$S_b=R_b=N_b=0$ and we note that the typical imbalance of populations
is $\sqrt{\mathrm{Var}(N'_a-N'_b)}=\sqrt{N_a}$.  Now, we use the
nontrivial state created by photodissociation as squeezed vacuum. We
imagine to have experimental control over $\theta$ and choose this
phase optimally in order to reduce $\mathrm{Var}(N'_a-N'_b)$.

\begin{figure}[htbp]
  \begin{center}
    \resizebox{8.5cm}{!}
    {
      \includegraphics{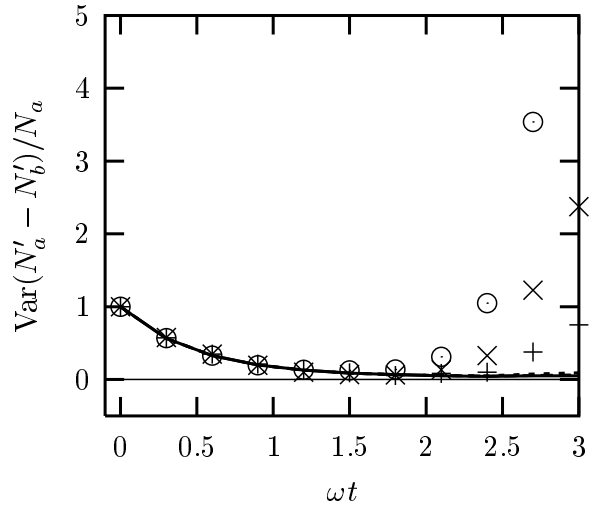}
      }
  \end{center}
  \caption{
    The minimum value of $\mathrm{Var}(N'_a-N'_b)/N_a$ as a function
    of the time of application of the $\pi/2$-pulse. All data are for
    a situation where incoupling is stopped at $\omega t = 2.4$.
    Calculations were done with three different interaction strengths
    ($g=0.005$, $0.01$ and $0.02$) and the almost overlapping lines
    (solid, dashed and dotted) show the results of R\&S equations
    while the symbols ($+$, $\times$ and $\odot$) show the results of
    the $P_+$ simulations.}
  \label{fig:numvar}
\end{figure}

In Fig.~\ref{fig:numvar} we plot this minimum value of
$\mathrm{Var}(N'_a-N'_b)/N_a$ as a function of the time of production
of the squeezed condensate. It is clearly seen how the noise is rather
quickly suppressed almost perfectly. In the particular case shown
$N_a$ was taken to be $10^3$ and $\phi_a$ was of gaussian shape.  The
time-dependent number of atoms in the $b$-condensate can be
approximately read out of Fig. \ref{fig:natoms}. The width of $\phi_a$
was chosen to be the same as the equilibrium width of the $b$-state
atoms.  It is amazing, but of course already well-known in quantum
optics, that this supression can take place even though the number of
atoms initially in the $b$-state is very small compared to the average
number of atoms in the large condensate in the $a$-state
  
\begin{figure}[htbp]
  \begin{center} 
     \resizebox{8.5cm}{!}
    {
      \includegraphics{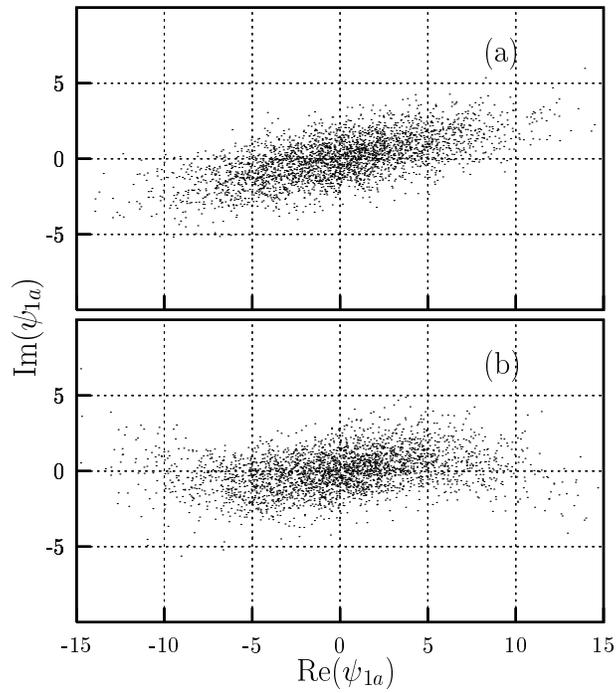}
    }
  \end{center} 
  \caption{Scatter plots of $\psi_{1a}$, the overlap of $\psi_1$ with
    the condensate mode function $\phi_a/\sqrt{N_a}$. Plot (a) is for
    $g=0.00$ while plot (b) is for $g=0.02$. The plots demonstrate the
    deforming effect of the interactions on the gaussian state. This
    deformation reduces the mean of $\psi_{1a}^2$ which is the
    decisive factor for the efficiency of the state as squeezed
    vacuum.}
  \label{fig:deform}
\end{figure}

At times beyond $\omega t \cong 2$ the noise suppression is lost in
the exact $P_+$ simulations. We recall that for $g=0$, the R\&S
equations are exact, and it is thus natural to ascribe the discrepancy
between the two methods to the interactions, which, by analogy with
the Kerr-effect in optics, cause a deformation of the gaussian state.
In Fig. \ref{fig:deform} we show a scatter plot of the amplitudes of
the projections, $\psi_{1a}$, of 3000 $P_+$ realizations of $\psi_1$
on $\phi_a$ at $\omega t = 2.4$ for $g=0.00$ and for $g=0.02$.
Such plots should be interpreted with care, as in general the
\emph{pair} $(\psi_1,\psi_2)$ is needed to calculate all
normal-ordered expectations. The Kerr-effect, however, influences the
$S$-function, and to calculate $\langle \hat\psi^2 \rangle$ we only
need to average $\psi_1^2$. In our case the relevant contributions to
this average can thus be depicted by a scatter plot of $\psi_{1a}$
alone.

We see in Fig.~\ref{fig:deform}a a gaussian state represented by
points which
form an ellipse-like structure. If the distibution had no prefered
direction in phase-space  $\langle
\hat\psi^2 \rangle=\overline{\psi_1^2}$ would vanish, but this is
clearly not the case for our squeezed state. In Fig.~\ref{fig:deform}b
the the interaction modifies
the phase accumulation of points with large values of
$\left|\psi_1\right|$ and deforms the distribution to an S-like
shape, and the mean value of $\psi_1^2$ is reduced. If the state is
centered around a finite field amplitude, the intensity dependent
phase shift transforms a circular distribution into a bean-shaped one,
and in that case the Kerr effect actually produces squeezing~\cite{reynaud}.

\section{Discussion}
\label{sec:conclusion}

In this paper we have studied the matter wave analogue of an optical
parametric oscillator. We have developed a description based on the
correlations of the atomic field and we have supplemented this by the
$P_+$ method. When interactions are taken into account both treatments were
limited to relatively short times and low numbers of atoms. We showed
how the state can be used as squeezed vacuum to improve the statistics
of experiments with larger condensates.

On the theoretical side, it is of course interesting to remedy the
breakdown of the R\&S method as we know that for large condensates an
even simpler description, the Gross-Pitaevskii equation, is very
successful. To bridge the gap between the two regimes which are both
describable in simple terms would be interesting and could be useful
also in other scenarios where special quantum states of a BEC are
created. In such work it is very likely that time dependent Monte
Carlo methods, based for example on the positive P distribution,
$P_+$, will play a significant role.

Let us finish this paper with a brief mentioning of other proposals
for the preparation of quantum correlated atomic condensates.  As
mentioned in the introduction, the collisional interaction has been
proposed as an agent for the preparation of quantum correlated states,
such as Schr\"odinger cat-like states \cite{cat}.  The states
prepared by our proposal are substantially less 'non-classical', but
provided the existence of a molecular condensate, we believe that the
experimental requirements of our proposal are more easy to meet.

It was suggested some time ago to prepare spin-squeezed states of 
non-degenerate atomic gasses and of trapped ions \cite{wineland_ueda},  
and concrete proposals were made, involving coupling of the atoms {\it via} 
their interaction with a quantized
field mode \cite{agarwal}, or their center-of-mass motion
\cite{ion_et_lattice}. More recently, it was suggested, and
experimentally proven possible, to induce
transitions between atomic states in an optically thick sample by
absorption of 
non-classical light, and thereby to transfer quantum  correlations  from the
light field to the atoms \cite{light_atom}. Ingredients of these
proposals become even more powerful with cavities around the samples
\cite{fleischhauer}. Very promising, recent ideas are based on quantum
non-demolition (QND) measurements of, {\it e.g.}, quadratures or populations
in the atomic sample by refraction of light beams. Such detection 
suffices to establish quantum correlations in the
sample, even though the detection is done with classical field
states \cite{QND}. 
One may simply use the state following the QND measurement
together with the known classical outcome of the measurement to
perform subsequent high-precision experiments, or one may
consider feed-back loops, in which the system is driven towards
a specific quantum correlated state.

All of these proposals are also relevant for
degenerate gasses. The QND methods are probably closest to 
real implementation, since phase contrast methods, already
used for imaging of condensates in many experiments, only have to 
be carried out with properly chosen parameters to lead to
quantum correlated atomic states. But all methods are 
interesting, and they may illuminate various aspects of the
condensate dynamics, as illustrated, {\it e.g.} by the 
threshold phenomenon and the 'Kerr-effect' suppression of
squeezing in the proposal in this paper.

\appendix

\section*{Synthesis of correlated noise}
\label{sec:correlated_noise}

In order to numerically simulate
Eqs.(\ref{eq:langevin1},\ref{eq:langevin2}) we discretize time and
space, and we 
synthesize noise terms, $dW_{1,2}(x_n,t_i)$, that obey discretized
versions of Eqs.(\ref{eq:wiener}-\ref{eq:wiener22}).  It is well-known how independent
(pseudo) random numbers from different distributions can be created.
For example a Gaussian distribution can be created starting from
uniformly distributed numbers {\it via} the Box-M{\"u}ller method,
{\it i.e.} we know how to produce $\{dU_{1,2}(x_n,t_i)\}$ so that
\begin{eqnarray}
  \label{eq:dU}
  \langle dU_{1,2}(x_n,t_i) \rangle &=& 0  \\
  \label{eq:dUdU}
  \langle dU_{\alpha}(x_n,t_i) dU_{\beta}(x_m,t_j) \rangle
  &=& \delta_{\alpha\beta}\delta_{nm}\delta_{ij}.
\end{eqnarray}
We see that the correlation functions,
Eqs.(\ref{eq:wiener11},\ref{eq:wiener22}), contain two terms: one
from the interaction and one from the incoupling. These can be treated
separately if we split the noise in two independent contributions
\begin{equation}
  \label{eq:split_dW}
  dW_{1,2}(x_n,t_i)=dW^{g}_{1,2}(x_n,t_i)+dW^{b}_{1,2}(x_n,t_i)
\end{equation}
Due to the contact form of the interaction, the corresponding noise
term poses no difficulties; we simply choose
\begin{equation}
  \label{eq:dW_g}
  dW^g_{1,2}(x_n,t_i)
  =\sqrt{\frac{\pm i g \psi(x_i) dt}{dx}}dU^g_{1,2}(x_n,t_i),
\end{equation}
where $dU^g_{1,2}$ is chosen with the properties (\ref{eq:dU},\ref{eq:dUdU}).  The
incoupling term is created by multiplication and convolution of
uncorrelated noise $dU^b_{1,2}$ of the form (\ref{eq:dU},\ref{eq:dUdU}) with 
suitable Gaussian functions:
\begin{equation}
  \label{eq:dW_b_ansatz}
  dW^b_{1,2}(x_n,t_i) = {\cal N} \exp \left( \pm i t_i \Delta \right) 
    \exp \left(
    - \frac{x_n^2}{2\sigma_a^2} 
      \right)
      \sum_{n'} dx \; dU^b_{1,2}(x_{n'},t_i) \exp \left(
    - \frac{(x_n-x_{n'})^2}{2\sigma_b^2 } 
    \right) 
\end{equation}
It turns out that choosing
\begin{equation}
    \label{eq:sigma_a_sigma_b_mathcal_N}
    \sigma_a^2=2\sigma_{cm}^2 ,
    \qquad
    \sigma_b^2=\frac{2\sigma_{cm}^2\sigma_r^2}{4\sigma_{cm}^2-\sigma_r^2} ,
    \qquad
    {\cal N}=\sqrt{\frac{\pm i dt 
 B}{dx\sqrt{\pi}\sigma_r\sigma_{cm}\sigma_b}}
\end{equation}
is sufficient to fulfill Eq.(\ref{eq:wiener}) with $b$ given by
Eq.(\ref{eq:b_def}). $\sigma_r$ and $\sigma_b$ are rather small, and in
practice the sum in Eq.(\ref{eq:dW_b_ansatz}) only needs to involve a
few terms.

\appendix

\clearpage

\end{document}